\documentclass[preprint,showpacs,preprintnumbers,amsmath,amssymb]{revtex4}

\usepackage{graphicx}
\usepackage{dcolumn}
\usepackage{bm}

\def\pt{\partial}

\def\al{\alpha}

\def\eps{\varepsilon}

\def\Om{\Omega}
\def\om{\omega}

\begin{document}

\title{Shock wave surfing acceleration}

\author{A. A. Vasiliev}
\email{valex@iki.rssi.ru}

\affiliation{Space Research Institute,\\
Profsoyuznaya str. 84/32, 117997 Moscow, Russia}

\begin{abstract}
Dynamics of a charged relativistic particle in a uniform magnetic field and an obliquely
propagating electrostatic shock wave is considered. The system is reduced to a two
degrees of freedom Hamiltonian system with slow and fast variables. In this system, the
phenomenon of capture into resonance can take place. Under certain condition, a captured
phase point stays captured forever. This corresponds to unlimited surfing acceleration of
the particle. The preprint is a more detailed version of a comment on the paper by D.Ucer
and V.D.Shapiro \cite{USh}, intended for the Comments section of Physical Reviews
Letters.
\end{abstract}


\maketitle

In Letter \cite{USh}, unlimited surfing acceleration of relativistic particles by a shock
wave normal to a uniform magnetic field was considered. The mechanisms of surfing
acceleration were studied in many papers, mainly in the case of acceleration by a
harmonic wave. The aim of this comment is to point out that the methods developed in
\cite{INV} allow for more detailed study and further progress in the topic of shock wave
acceleration. In particular, the case of an oblique shock wave can be considered. Also,
it can be shown that particles with initial velocities far from the shock wave velocity
can also be captured in the mode of unlimited acceleration.

Like in \cite{USh}, consider a charged relativistic particle of charge $e$ and rest mass
$m$ in a uniform magnetic field ${\bf B}$ and an electrostatic shock wave of potential
$\Phi = - \Phi_0 \tanh({\bf kq} -\om t)$, where $\Phi_0 >0, \, \om >0, \, \om/|{\bf
k}|=u$ is the phase velocity of the shock wave, ${\bf q}$ is the radius vector. Choose an
orthogonal coordinate system $(q_1,q_2,q_3)$ such that ${\bf B}=B_0 {\bf e_3}$ is along
the $q_3$-axis and ${\bf k}$ lies in the $(q_1,q_3)$-plane, ${\bf k}=(k_1,0,k_3)$. The
Hamiltonian function of the particle is
\begin{equation}
H=(m^2 c^4 +c^2 p_1^2 +c^2 p_3^2 +(c{\cal P}_2 - e B_0 q_1)^2)^{1/2} - e\Phi_0
\tanh(k_1 q_1 +k_3 q_3 -\om t), \label{1}
\end{equation}
where ${\cal P}_2= p_2+ eB_0 q_1/c$ and ${\bf p}=(p_1,p_2,p_3)$ is the particle's
momentum. Introduce notations:
\begin{equation}
\om_c= \frac{eB_0}{mc},\; k=(k_1^2 +k_3^2)^{1/2},\; \eps=\frac{e\Phi_0}{mc^2},\; \Om_c=
\om_c/\eps,\; \sin\al = k_3/k. \nonumber
\end{equation}
Consider the problem
in the following range of parameters: $|{\bf p}|/(mc) \sim 1,\, \om/(kc) \sim 1,\, \eps
\ll 1,\, \om_c/\om \sim \eps$. Rescale the variables: $\tilde{p}_{1,3}=p_{1,3}/(mc),\,
\tilde{q}_{1,3}=\eps q_{1,3}/c,\, \tilde{k}_{1,3}=k_{1,3}c,\, \tilde{H}=H/(mc^2)$.
Following \cite{INV}, one can canonically transform (\ref{1}) into the form (tildes are
omitted):
\begin{equation}
{\cal H} = -\om I +\left[1 +k^2(I +p\cos\al/k)^2 +p^2\sin^2\al +\Om_c^2 q^2
\right]^{1/2} -\eps\tanh\phi \equiv {\cal H}_0 -\eps\tanh\phi, \label{2}
\end{equation}
where canonically conjugated pairs of variables are $(p, \eps^{-1}q)$ and $(I,\phi)$,
$\phi =k_1 q_1 +k_3 q_3 -\om t$. The corresponding Hamiltonian equations of motion imply
that while $\dot{\phi}\neq 0$, total change in variable $I$ is a value of order $\eps$,
and hence the trajectory of the particle in the $(p,q,I)$-space lies in a vicinity of the
intersection of a second order surface ${\cal H}_0 =$const and a plane $I=$const. This
intersection is an ellipse corresponding to the Larmor motion. However, along a
trajectory that crosses the resonance $\dot{\phi}=\pt{\cal H}_0/\pt I=0$ the value of $I$
can change significantly.

\begin{figure*}
\includegraphics{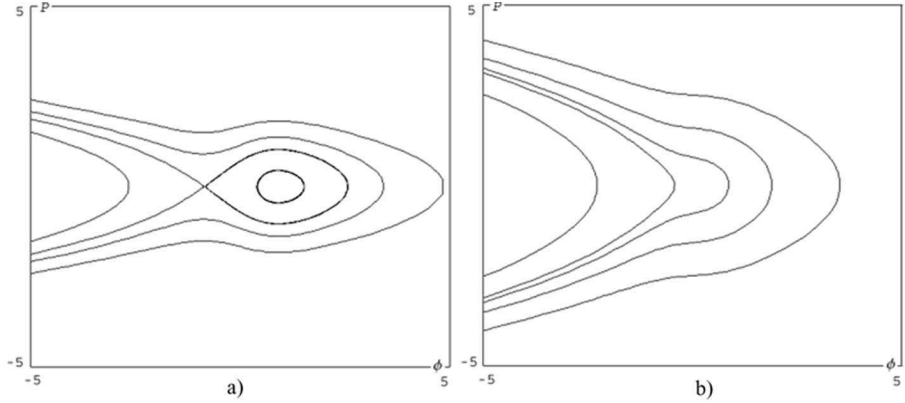}
\caption{\label{pendlike} Phase portraits of the ``pendulum-like'' system. a) $b=0.5$, b)
$b=1.2$.}
\end{figure*}

The resonant condition $\pt{\cal H}_0/\pt I=0$ defines a surface $I=I_{res}(p,q)$ in the
$(p,q,I)$-space, called the resonant surface. The condition implies that projection of
the particle's velocity onto the direction of vector ${\bf k}$ equals the phase velocity
of the wave. Intersection of the resonant surface and the surface ${\cal H}_0=$const is a
second order curve whose kind depends on the parameter values (see \cite{INV}). This
curve is called the resonant curve. The motion in a neighborhood (of the width of order
$\sqrt\eps$) of the resonant surface possesses certain universal properties (\cite{INV},
\cite{N98}). In particular, the Hamiltonian $F={\cal H}/\eps$ of the particle in this
neighborhood can be written in the form:
\begin{equation}
F=\eps^{-1} \Lambda(p,q) + F_0(P,\phi,p,q) +O(\sqrt\eps), \label{3}
\end{equation}
where $\Lambda(p,q)$ is ${\cal H}_0$ restricted onto the resonant surface,
$P=(I-I_{res}(p,q))/\sqrt\eps +O(\sqrt\eps)=O(1)$, and canonically conjugated pairs of
variables are $(P,\phi)$ and $(p,\eps^{-3/2}q)$. The function $F_0$ is so-called
``pendulum-like'' Hamiltonian, and in the case under consideration it is $F_0=
g(p,q)P^2/2 - \tanh\phi +b(p,q)\phi$, where
\begin{equation}
b(p,q)=\frac{\Om_c^2 \cos\al}{(k^2 -\om^2)^{1/2}}\cdot \frac{q}{(1+ p^2\sin^2\al
+\Om_c^2 q^2)^{1/2}},\qquad g(p,q)=\frac{k^2(1-(\om/k)^2)^{3/2}}{(1+ p^2\sin^2\al
+\Om_c^2 q^2)^{1/2}}, \label{4}
\end{equation}
In the system defined by $F$, variables $(p,q)$ are slow and variables $(P,\phi)$ are
fast. Slow evolution of $(p,q)$ is determined by a system with Hamiltonian $\sqrt\eps
\Lambda$. This system defines a flow on the resonant surface, called resonant flow. The
$(P,\phi)$ variables evolve according to the subsystem with Hamiltonian $F_0$. If
$0<b<1$, there is a separatrix surrounding the oscillation region on the phase portrait
of this subsystem (see Figure \ref{pendlike}). If $b<0$ or $b>1$, there is no oscillation
region.

The area of the oscillation region $S$ is a function of the slow variables: $S=S(p,q)$.
If $S(p,q)$ enlarges along the resonant flow, additional area appears inside the
oscillation region. Hence, phase points cross the separatrix and enter the oscillation
region. This is a capture into resonance. A captured phase point leaves a vicinity of the
curve $I=$const, ${\cal H}_0=$const and continues its motion following approximately the
resonant curve. Note, that phase points with arbitrarily large initial values of $P$ can
be captured provided they are close enough to the incoming invariant manifold of the
saddle point of the ``pendulum-like'' system. This corresponds to the fact that a
particle can be trapped in the mode of surfing acceleration even in the case that
initially it is far from the resonance.

The area bounded by the trajectory of a captured phase point in the $(P,\phi)$-plane is
an adiabatic invariant of the ``pendulum-like'' system. Hence, if $S(p,q)$ contracts
along the resonant flow, some phase points leave the oscillation region and leave the
resonant zone. This is an escape from the resonance. If $S$ monotonically grows along the
resonant curve, none of phase points leave the oscillation region. In this case, captured
phase points stay captured forever.

If the resonant curve is a hyperbola ($k_3<\om$, \cite{INV}) or a parabola ($k_3=\om,
{\cal H}_0<0$, \cite{INV}), a captured phase point may go to infinity. In this motion the
energy of the particle $H$ (see (\ref{1})) tends to infinity. Therefore, this motion
produces unlimited surfing acceleration of particles. This acceleration is possible, if
$S(p,q)$ grows as $p,q \to \infty$ along the resonant curve. Calculations (see
\cite{INV}) give the following necessary condition of possibility of unlimited
acceleration:
\begin{equation}
\frac{\Om_c(\om^2 -k^2\sin^2\al)^{1/2}}{\om(k^2- \om^2)^{1/2}}< 1. \label{5}
\end{equation}
This condition was first obtained in \cite{ChScN} for acceleration by a harmonic wave. In
the case of perpendicular propagation, it is equivalent to the condition of Katsouleas
and Dawson \cite{KatsDaw}, also mentioned in \cite{USh}.

Consider a hyperbolic resonant curve under assumption that (\ref{5}) is valid. At $q<0$,
$b(p,q)<0$ (see (\ref{4})) and $S(p,q)=0$. At $q=0$, function $S(p,q)$ has a singularity
and at small positive $q$ it is very large. As $q$ grows along the resonant curve,
$S(p,q)$ first decreases and then, as $q \to \infty$, $S(p,q) \to \infty$. Hence, at a
certain $q=q_m$, function $S(p,q)$ has minimum $S=S_m$ along the resonant curve. Consider
a phase point that is initially captured into the resonance at small positive value of
$q$. Let the area bounded by its trajectory be $S_0$ and $S_0>S_m$. Then in the course of
motion along the resonant curve this phase point escapes from the resonance. If
$S_0<S_m$, the phase point stays captured forever and undergoes unlimited acceleration.
This explains Fig. 5 in \cite{USh}. The number of bounces in this figure corresponds to
the number of oscillations of the phase point inside the oscillatory region of the
``pendulum-like'' system, performed before the phase point escapes from the resonance.

The author thanks A.I.Neishtadt for useful discussions. The work was partially supported
by grants RFBR 00-01-00538 and INTAS 00-221.


\begin{thebibliography}{99}

\bibitem{USh} D.Ucer and V.D.Shapiro, Unlimited Relativistic Shock Surfing Acceleration,
Phys.Rev.Lett. {\bf 87}, 075001 (2001)

\bibitem{INV} A.P.Itin, A.I.Neishtadt, and A.A.Vasiliev, Captures
into resonance and scattering on resonance in dynamics of a charged relativistic particle
in magnetic field and electrostatic wave, Physica D {\bf 141} (2000) 281-296.

\bibitem{N98} A.I.Neishtadt, On Adiabatic Invariance in Two-Frequency
Systems, In: "Hamiltonian systems with three or more degrees of freedom", Ed. C.Simo,
NATO ASI Series, Series C, vol. 533, Kluwer Academic Publishers, Dordrecht/Boston/London,
1999, 193-213.

\bibitem{ChScN} A.A.Chernikov, G.Schmidt, and A.I.Neishtadt, Unlimited
particle acceleration by waves in a magnetic field,  Phys.Rev.Letters {\bf 68} (1992)
1507-1510.

\bibitem{KatsDaw} T.Katsouleas and J.M.Dawson, Unlimited Electron
Acceleration in Laser-Driven Plasma Waves, Phys.Rev.Lett. {\bf 51} (1985) 392-395.


\end{thebibliography}
\end{document}